\documentclass[twocolumn]{aastex61}
\received{July 1, 2016}
\revised{September 27, 2016}
\accepted{\today}
\submitjournal{PASP}

\shorttitle{Transient Universe in Ultraviolet}
\shortauthors{Wang et al.}

\begin{document}

\title{Learning Transient Universe in  Near-Ultraviolet  By Wide-angle Cameras}

\correspondingauthor{J. Wang}
\email{wj@bao.ac.cn; wj@gxu.edu.cn}

\author{J. Wang}
\affil{Guangxi Key Laboratory for Relativistic Astrophysics, School
of Physical Science and Technology, Guangxi University, Nanning
530004, People's Republic of China}
 \affil{Key Laboratory of Space
Astronomy and Technology, National Astronomical Observatories,
Chinese Academy of Sciences, Beijing 100101, China}
 \affil{School of
Astronomy and Space Science, University of Chinese Academy of
Sciences, Beijing, China}

\author{E. W. Liang}
\affil{Guangxi Key Laboratory for Relativistic Astrophysics, School
of Physical Science and Technology, Guangxi University, Nanning
530004, People's Republic of China}

\author{J. Y. Wei}
\affiliation{Key Laboratory of Space Astronomy and Technology, National Astronomical Observatories, Chinese Academy of Sciences, Beijing
100101, China}
\affiliation{School of Astronomy and Space Science, University of Chinese Academy of Sciences, Beijing, China}



\begin{abstract}

We perform a detailed analysis and simulations on the transient
detection capability in the near-ultraviolet (NUV) band by focusing
on some major local transient events. These events include the tidal
disruption event due to a supermassive blackhole, the shock breakout
of a core-collapse supernova and the flare of a late-type star. Our
simulations show that a set of small wide-angle NUV cameras  can
allow us to detect and study numerous galactic and extra-galactic
transient events. Based on the analysis and simulations, here we
propose a space-based NUV sky patrol mission by updating the
proposal that was originally submitted to the Chinese Space Station
mission in 2011. The mission proposed here is composed of a set of
eight small wide-field NUV cameras each with a diameter of 20cm. The
total sky area simultaneously covered by the NUV cameras is as large
as 3000$\mathrm{deg^2}$. The survey cadence ranges from 30 to 300s.
The transient events are required to be detected by a dedicated
on-board software in real time.
\end{abstract}




\section{Introduction} \label{sec:intro}

Time-domain astronomy, which mainly explores the Universe by
searching for and studying transients in different wavelength bands,
is a hot topic of modern astronomy.  Transient phenomena have been
widely explored in high energy bands (i.e., X-ray and $\gamma$-ray)
in the past decades, thanks to the great success achieved by the
Neil Gehrels \it Swift \rm Observatory (Gehrels et al. 2004) and the
\it Fermi\rm\ Gamma-ray Space Telescope (e.g., Atwood et al. 2009;
Meegan et al. 2009). This success enables us to make great progress
in the understanding of Gamma-ray bursts (GRBs, see a review in
Woosley \& Bloom 2006). The capability of exploring transients in
X-ray and $\gamma$-ray will be continued by forthcoming missions,
including SVOM (Wei et al. 2016) and the Einstein Probe (EP) (Yuan
et al. 2015).

In addition to the high energy bands, transient phenomena have been
comprehensively probed by a lot of  optical time-domain surveys in
recent years, because of the great progress made in both detector
and computation technology in the past decades. Some ongoing optical
survey programs include: the Pan-STARRS1  survey (PS1; Chambers et
al. 2016), the Asteroids Terrestrial-impact Last Alert System
(ATLAS; Tonry et al. 2018), the Palomar Transient Factory (PTF; Law
et al. 2009) and the successor Zwicky Transient Facility (ZTF;
Kulkarni 2018), the All-Sky Automated Survey for Supernovae
(ASAS-SN; Shappee et al. 2014), Pi of the sky (Burd et al. 2005),
RAPTOR (Vestrand et al. 2002), and the Ground Wide-Angle Cameras
system (GWAC, Wei et al. 2016). These optical transient surveys
discovered an abundance of supernovae (SNe) in diverse types and the
related phenomena (e.g., Gal-Yam et al. 2014; Arcavi et al. 2017;
Whitesides et al. 2017), a number of tidal disruption events (TDEs)
induced by supermassive blackholes (SMBHs, e.g., Holoien et al. 2018
and reference therein), ``Changing-look'' active galactic nuclei
(CL-AGNs, e.g., Gezari et al. 2017; Wang et al. 2018), and some
mysterious fast-evolving luminous transients without a convincing
explanation (e.g., Rest et al. 2018 and references therein). The
forthcoming Large Synoptic Survey Telescope (LSST) project (LSST
Science Collaboration et al. 2017) will be a milestone for optical
transient survey. The transient sky has been monitored by LOFAR
(e.g., Gunst \& Bentum 2007; de Vos et al. 2007) and the SKA in
radio.

In addition to electromagnetic radiation,  a non-electromagnetic
message transmitted by  a gravitational-wave  (e.g., the Advanced
LIGO, Harry 2010; Aasi et al. 2015; LIGO Scientific Collaboration et
al., 2015, the Advanced Virgo, Acernese et al. 2009; Accadia et al.
2012; Acernese et al. 2015; KAGRA, Somiya 2012; Aso et al. 2013) and
by neutrino events (e.g., the IceCube experiment, Aartsen et al.
2015) has been used to explore transient phenomena. Operation of
Advanced LIGO opened an era of multi-message astronomy by
successfully detecting the gravitational-wave signals emitted from
the coalescence of a binary black holes (BBHs, e.g., GW\,150914,
Abbott et al. 2016) and the coalescence of a binary neutron stars
(BNS, GW\,170817. e.g., Abbott et al. 2017). The coalescence of
binary neutron stars was confirmed by the detection of
electromagnetic emission from the associated kilonova (e.g., Shappee
et al. 2017; Drout et al. 2017; Evans et al. 2017).

Time-domain astronomy is, however, rarely developed in ultraviolet
(UV) bands, even though the study of transients in UV can
potentially address some major scientific questions (e.g., Sagiv et
al. 2014; Brosch et al. 2014).  In fact, observations carried out by
\it Swift\rm/UVOT \rm and \it GALEX \rm (Martin et al. 2005)
returned exciting results on transients (e.g., Soderberg et al.
2008; Schawinski et al. 2008), although both instruments are not
designed to search for transients based on a large field-of-view
(FoV). The main scientific goal of \it GALEX \rm is to study star
formation and galaxy evolution in UV bands, which results in very
limited time-domain observations.

In the development of the Chinese Space Station mission, we proposed
a wide-angle time domain survey in near-UV (NUV) as early as in
2011. Subsequent studies, however, indicated that the Space Station
is not an ideal platform for this survey. The depth of the survey is
limited because the on-board cameras can only work in drift scan
mode.

In this paper, we perform a detailed analysis on the detection
capability of an NUV wide-angle time domain survey. The analysis
focuses on some important scientific objects.  One can see from our
analysis that an abundance of results can be obtained by a set of
small NUV cameras. Based on this analysis, we then propose a small
satellite mission dedicated to the NUV wide-angle time domain
survey.

The paper is organized as follows. Section 2 describes the major
scientific motivations for the proposed NUV transient survey by
supplementing the  CL-AGN phenomenon. The analysis and simulation of
detection capability  are presented in Section 3. Section 4
describes the concept  of  the proposed dedicated NUV transient
survey mission. A $\Lambda$CDM cosmology with parameters
$H_0=70\mathrm{km\ s^{-1}\ Mpc^{-1}}$, $\Omega_m=0.3$, and
$\Omega_\Lambda=0.7$ is adopted throughout the paper

\section{Scientific Objectives} \label{sec:style}

For transient phenomena that can be preferably explored in the UV
band, we refer readers to the excellent review given by Sagiv et al.
(2014, and references therein). The phenomena described in that
review include: the supernova shock breakout (SBO) resulting from
the explosive death of a massive star; the TDE of a main-sequence
star (or a white dwarf) due to the tidal force of a supermassive (or
intermediate massive) blackhole; the variability of AGNs; and the
flare of stars.

Besides the topics described in the review, here we pay more
attention to the CL-AGN phenomenon from a scientific perspective,
because the phenomenon is a hot and challenging topic in modern
astronomy. According to their observed optical spectra, AGNs can be
classified into Type-1 and Type-2 AGNs. The spectra of Type-1 AGNs
have both broad ($\mathrm{FWHM>1000\ km\ s^{-1}}$) and narrow
($\mathrm{FWHM\sim10^2\ km\ s^{-1}}$) Balmer emission lines. On the
contrary, only narrow Balmer emission lines can be identified in
Type-2 AGNs. The two types can be unified by the widely accepted
unified model based on the orientation effect caused by the dust
torus (see Antonucci 1993 for a review). This successful model has,
however, been recently challenged by the discovery of so-called
CL-AGNs that show a change in their spectral types on a time scale
of several years.

Although both ``turn-on'' and ``turn-off'' type transitions have
been revealed in past years, there are only 40 identified CL-AGNs at
present (e.g., Shapovalova et al. 2010; Shappee et al. 2014; LaMassa
et al. 2015; McElroy et al. 2016; Runnoe et al. 2016; Gezari et al.
2017; Yang et al. 2018; Ruan et al. 2016; MacLeod et al. 2016; Wang
et al. 2018). The discovery of CL-AGNs is actually a hard and
expensive task, which requires repeated spectroscopy to identify a
change in the spectral type of the Balmer line profiles. At the same
time, an optical transient survey is not a good way to select CL-AGN
candidates because of the serious contamination caused by AGN's normal variation
in optical bands.

The origin of CL-AGNs is still under debate, even though there is
accumulating evidence supporting the fact that it is likely due to a
variation in SMBH accretion rate that results from either a viscous
radial inflow or disk instability (e.g., Yang et al. 2018; Wang et
al. 2018; Gezari et al. 2017). In addition, other explanations
include: (1) a variation in the obscuration if the torus has a
patchy configuration (e.g., Elitzur 2012); (2) an accelerating
outflow launched from the central SMBH (e.g., Shapovalova et al.
2010); and (3) a TDE (e.g., Merloni et al. 2015; Blanchard et al.
2017).

Besides the prominent line profile change, CL-AGNs are typically
accompanied by a significant variation in their blue featureless
continuum, because Type-1 and Type-2 AGNs differ significantly in
their UV continuum. This significant difference in continuum means
an NUV transient survey is the best way to find both ``turn-on'' and
``turn-off''  CL-AGNs in the local Universe.

\section{Detection Rate Predictions} \label{subsec:tables}

For a wide-angle time domain survey in NUV,  the detection rates of
some important scientific objects are predicted in this section.


\subsection{UV Brightness Estimated from X-ray Flux}

At the beginning, we estimate the NUV brightness of some transients
from their soft X-ray flux.   A  powerlaw photon spectrum of $N(E)=N_0E^{-\Gamma}$
is adopted for GRBs, SBOs, and TDEs. The
corresponding X-ray flux $f_\mathrm{X}$ within the energy range from
$E_1$ to $E_2$ can be written as an integral of
$f_\mathrm{X}=\int_{E_1}^{E_2}N(E)EdE=N_0\int_{E_1}^{E_2}
E^{1-\Gamma}dE$, which allows us to obtain the specific flux $f_\nu$ at a given wavelength
$\lambda$ as 
\begin{equation}
  f_\nu=EN(E)\frac{dE}{d\nu}=\frac{\lambda f_\mathrm{X}}{c}\bigg[\frac{2-\Gamma}{(E_2/E)^{2-\Gamma}-(E_1/E)^{2-\Gamma}}\bigg]
 \end{equation}
when $\Gamma\neq2$, and
\begin{equation}
  f_\nu=EN(E)\frac{dE}{d\nu}=\frac{\lambda f_\mathrm{X}}{c}\ln^{-1}\bigg(\frac{E_2}{E_1}\bigg)
 \end{equation}
 when $\Gamma=2$.  Based on the traditionally used Band function (Band et al. 1993), the values of $\Gamma$ are adopted to be 1.0 and 0.5 for Long GRBs and short GRBs, according to the BASTE and Fermi observations (e.g., 
 Abdo et al. 2009; Zhang et al. 2011, Nova et al. 2011).  A soft spectrum with a $\Gamma=2.0$ is used for 
 SBOs and TDEs.   In fact, the XMM-Newton 
 spectrum of TDE RXJ1242¨C1119 has an index of $\Gamma=2.5$, although the initial spectrum taken by ROSAT has an index of $\Gamma\simeq5$ (e.g., Komossa 2017; Komossa et al. 2004; 
 Halpern et al. 2004). 
 In fact, similar spectral hardening has been observed in a few TDEs (e.g., Komossa
\& Bade 1999; Nikolajuk \& Walter 2013). Soderberg et al. (2008) revealed a X-ray spectrum with 
a photon index of $\Gamma=2.3\pm0.3$ in SBO event of SN\,2008D.

We estimate the UV brightness of a stellar flare from its soft X-ray flux according to the Neupert effect 
which suggests a correlation between soft X-ray luminosity and UV energy release 
(e.g., Gudel et al. 2002; Hawley et al. 1995). A average ratio of X-ray to NUV specific luminosity of 
20 is adopted in our estimation.

The corresponding magnitude in the AB system is then determined from the definition
(Fukugita et al. 1996)
\begin{equation}
 \mathrm{mag_{AB}}=-2.5\log f_\nu-48.6+A_\lambda
\end{equation}
where $A_\lambda$ denotes the extinction at wavelength $\lambda$.
Table 1 lists the calculated magnitudes at different X-ray flux levels, in which $\lambda=2800$\AA, $A_\lambda(2800\AA)=0.2$mag, $E_1=0.5$keV and $E_2=4.0$keV
are adopted. The $A_\lambda(2800\AA)$ is estimated from the $V$-band extinction through the extinction curve of LMC provided
in Gordon et al. (2003), by assuming $A_V=0.1$mag.

\begin{table}[h!]
\renewcommand{\thetable}{\arabic{table}}
\centering
\caption{UV brightness estimated from different soft X-ray fluxes.}
\label{tab:decimal}
\begin{tabular}{cccc}
\tablewidth{0pt}
\hline
\hline
$f_x$ &  $\mathrm{mag_{AB}}$ & $\Gamma$  & Transients\\
  $\mathrm{erg\ s^{-1}\ cm^{-2}}$   &  & mag &  \\
(1)  &   (2) & (3) & (4) \\
\hline
$10^{-7}$  &  14.3 & 1.0 & LGRB\\
$10^{-8}$  &  17.7 &  1.0 & LLGRB\\
                  &  20.0 &  0.5 & SGRB\\
$10^{-9}$  &  19.3 &  1.0 & LLGRB\\
                  &  12.6 & 2.0   & SBO\\
$10^{-10}$  & 15.1&  2.0 & SBO, TDE\\
                    & 9.8 & \dotfill & stellar flare\\
$10^{-11}$  & 17.6 & 2.0 & SBO,TDE\\
                    & 12.3 & \dotfill & stellar flare\\

\hline
\hline
\end{tabular}
\tablecomments{LGRB: Long gamma-ray burst; LLGRB: Low-luminosity gamma-ray burst; 
SGRB: Short gamma-ray burst; SBO: Shock breakout; TDE: Tidal disruption event. }
\end{table}

\subsection{Comparison of Spectral Energy Distributions}

Compared to surveys in optical, one of the advantages of an NUV
time-domain survey is the extremely low brightness of the host
galaxies of extra-galactic transients. To clarify this fact, here we
compare the SEDs of TDEs and  of SBOs to those of their hosts. It is
emphasized that the host galaxies are hard to spatially resolve in a
wide-angle survey, except for very nearby galaxies.

\subsubsection{TDEs}

For a given SMBH with a mass of $M_{\mathrm{BH}}$, the SED of a TDE
of a main-sequence star with mass $M_*$ is modeled by following
Lodato \& Rossi (2011) by including the contributions from both a
hot accretion disk and a disk wind.

When the effect due to the disk wind is considered, the disk
temperature $T_d$ can be written as a function of the distance $R$
from the central SMBH as (Strubbe \& Quataert 2009)
\begin{equation}
  \sigma T_d^4=\frac{3GM_{\mathrm{BH}}\dot{M}f}{8\pi R^3}\bigg\{\frac{1}{2}+\bigg[\frac{1}{4}+\frac{3}{2}f\bigg(\frac{\dot{M}}{\eta \dot{M}_{\mathrm{Edd}}}\bigg)^2
  \bigg(\frac{R_\mathrm{S}}{R}\bigg)^2\bigg]^{1/2}\bigg\}^{-1}
\end{equation}
where $R_\mathrm{S}=2GM_{\mathrm{BH}}/c^2$ is the Schwartzchild
radius of the SMBH,
$\dot{M}_\mathrm{Edd}=1.3\times10^{18}(M_{\mathrm{BH}}/M_\odot)(\eta/0.1)^{-1}\
\mathrm{g\ s^{-1}}$ is the Eddington accretion rate, and
$f=1-\sqrt{R_{\mathrm{in}}/R}$. We adopt the innermost stable
circular orbit of the disk $R_{\mathrm{in}}=3R_\mathrm{S}$
throughout the subsequent analysis. The accretion rate $\dot{M}$ is
related to the fallback rate $\dot{M}_{\mathrm{fb}}$ as
$\dot{M}=(1-f_{\mathrm{out}})\dot{M}_{\mathrm{fb}}$. The factor
$f_{\mathrm{out}}$ accounts for the outflow effect due to the disk
wind, which can be expressed as (Ditan \& Shviv 2010)
\begin{equation}
  f_{\mathrm{out}}=\frac{2}{\pi}\arctan\bigg[\frac{1}{7.5}\bigg(\frac{\dot{M}_{\mathrm{fb}}}{\dot{M}_{\mathrm{Edd}}}-1\bigg)\bigg]
\end{equation}
for super Eddington accretion. $\dot{M}_{\mathrm{fb}}\propto
t^{-5/3}$, which has a peak value of
\begin{equation}
 \dot{M}_{\mathrm{fb,p}}=1.9\times10^{26}M_{\mathrm{BH,6}}^{-1/2}m_*^2\beta^3x_*^{-3/2} \mathrm{g\ s^{-1}}
\end{equation}
where $M_{\mathrm{BH,6}}=M_{\mathrm{BH,}}/10^6M_\odot$,
$m_*=M_*/M_\odot$, $x_*=R_*/R_\odot$, and $\beta=r_t/r_p$.
$r_t=10^{-2}M_{\mathrm{BH,6}}^{1/3}m_*^{-1/3}x_* R_\odot$ is the
tidal disruption radius and $r_p$ the periastron radius.

Because the emission from the hot accretion disk is dominant in soft
X-ray and UV, the photosphere model was proposed to explain the
optical emission detected in some TDEs (e.g., Guillochon et al.
2014; Roth et al. 2015; Jiang et al. 2016; Metzger \& Stone 2016).
Based on basic energy conservation, the radius and temperature of
the photosphere at the peak time are (Lodato \& Rossi 2011)
\begin{equation}
  r_{\mathrm{ph}}\approx1.4\times10^{15}\bigg(\frac{f_\mathrm{out}}{f_\upsilon}\bigg)\beta^{5/2}M_{\mathrm{BH,6}}^{-5/6}x_*^{-1}
  m_*^{11/6}\ \mathrm{cm}
\end{equation}

\begin{equation}
  T_{\mathrm{ph}}\approx1.6\times10^{4}\bigg(\frac{f_\mathrm{out}}{f_\upsilon}\bigg)^{-1/12}\beta^{-23/24}M_{\mathrm{BH,6}}^{-41/73}x_*^{1/3}
  m_*^{-53/72}\ \mathrm{K}
\end{equation}
where $f_\upsilon\geq1$,  in the unit of escape velocity, is a
parameter describing the wind velocity.

The total emission is therefore the combination of two components
\begin{equation}
  L_\nu=4\pi^2\int_{R_{\mathrm{in}}}^{R_{\mathrm{out}}}RB_\nu(T_d)dR+4\pi^2r_{\mathrm{ph}}^2B_\nu(T_{\mathrm{ph}})
\end{equation}
where $B_\nu(T)$ is the Planck function and
$R_{\mathrm{out}}\approx15\beta^{-1}M_6^{-1/2}m_*^{-1/3}x_*R_{\mathrm{in}}$
is the outer distance of the disk .

Figure 1 compares the predicted UV-to-optical SEDs of the TDEs to
the modeled SEDs of the corresponding host galaxies. The four panels
correspond to four cases with different $M_{\mathrm{BH}}$. The
following fiducial values are adopted in the predictions:
$m_*=x_*=\beta=1$, $\eta=0.1$, and $f_\upsilon=1$. An UV-to-optical
spectrum with an age of 1.4Gyr and a metalicity of 0.05$Z_\odot$ is
first extracted from the single stellar population (SSP) spectral
library built by Bruzual \& Charlot (2003). The luminosity level of
the spectrum is then scaled to the  bulge of the corresponding host
galaxy through the firmly established
$M_{\mathrm{BH}}-L_{\mathrm{bulge}}$ relationship: $\log
M_{\mathrm{BH}}=9.23+1.11\log(L_{\mathrm{bulge}}/10^{11}L_\odot)$
(McConnell \& Ma 2013). Two facts can be learned from the comparison
shown in the Figure 1. On the one hand, the TDEs are brighter than
their host galaxies by 1-3 orders of magnitudes in the FUV band in
all the four cases. The NUV band is quite sensitive to the TDEs with 
a blackhole mass of $10^6M_\odot$ by a brightness excess of about 3 orders of magnitudes, and is also 
plausible for a case with a  blackhole mass either $\sim10^5 M_\odot$ or 
$\sim10^7 M_\odot$. 
On the other hand, it is a hard task to detect a
TDE in optical bands in the case with an SMBH more massive than
$10^6M_\odot$.

\begin{figure}
\plotone{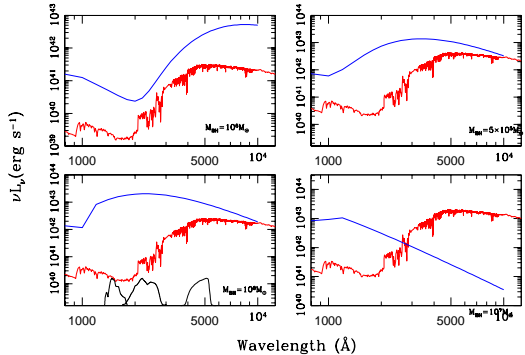} \caption{In each panel, the predicted
TDE's SED (the blue curve) is compared to the modeled SED of the
corresponding bulge (the red curve).  Different panels correspond to
different SMBH mass. The transmission curves of the FUV, NUV, and
SDSS $g$ band are shown in the left-bottom panel by black curves. }
\end{figure}

\begin{table}[h!]
\renewcommand{\thetable}{\arabic{table}}
\centering
\caption{The parameters used for detection rate estimations.}
\label{tab:decimal}
\begin{tabular}{cc}
\tablewidth{0pt}
\hline
\hline
Item & Value \\
(1)  &   (2) \\
\hline
$d$  &  20cm\\
$\eta$  &  0.3\\
RN  &  $4e^-$\\
CCD pixels & 4k$\times$4k\\
$n_{\mathrm{pix}}$ & 9 pixels\\
$\lambda_0$ & 2000\AA\\
$\Delta\lambda$ & 500\AA\\
Exposure time  & TDEs: 3000s, SBOs: 300s and flares: 30s\\
$\mathrm{S/N}$ threshold  &  7\\
\hline
\hline
\end{tabular}
\end{table}

\subsubsection{SBO}

The spectrum of both an SBO and the following shock cooling can be
described well as a blackbody with a temperature of $10^{4-5}$K
(e.g., Matzner \& McKee 1999; Sapir \& Halbertal 2014). Figure 2
compares the spectrum of a blackbody at a temperature of
$1\times10^5$K to those of the galaxies with different total masses
ranging from $10^9$ to $10^{11}M_\odot$. The specific luminosity of
the blackbody spectrum is determined by requiring its NUV flux to
equal the observed flux of Type II-P SN SNLS-04D3dc (Schawinski et
al. 2008; Gezari et al. 2008) at $z=0.1854$, i.e.,
$f_\nu=5\times10^{-29}\ \mathrm{erg\ s^{-1}\ cm^{-2}\ Hz^{-1}}$. The
host galaxy spectra used for comparison have an age of 1.4Gyr, which
are extracted from the SSP spectra library again. The total light of
the galaxies is involved in the comparison by taking into account
the poor spatial resolution of a wide-angle survey. Similar to the
case of TDE, the comparison shows that NUV the time-domain survey is
more powerful for detecting SBOs than optical surveys.

\begin{figure}
\plotone{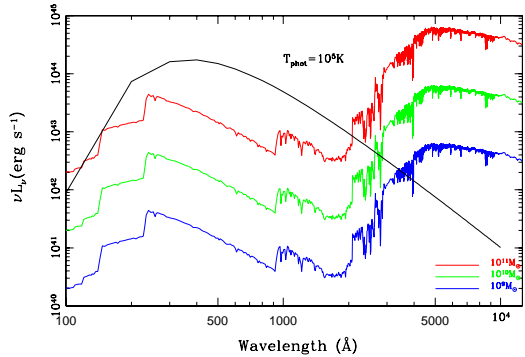} \caption{The SED (the black curve) of an
SBO with a temperature of $10^5$k is compared to the SED of the
total light of host galaxies with different total masses. }
\end{figure}

\subsection{Detection Rates}

After showing that contamination due to the underlying host galaxies
is negligible in an NUV survey, we then predict the detection rates
for TDEs, SBOs, and superflares of G- and late-type stars, for a
wide-angle time-domain survey in NUV. The survey is assumed to be
carried out by a set of wide-angle cameras that monitor different
sky areas simultaneously. Each camera is assumed to be equipped with
a 4k$\times$4k CCD as the detector.

We first determine the limiting magnitude of individual camera from
the simplified ``CCD'' equation (Merline \& Howell 1995; Mortara \&
Fowler 1981; Gullixson, 1992 and cf. NOAO/KPNO CCD instrument
manuals)
\begin{equation}
  \frac{S}{N}\simeq\frac{\eta N_*t}{\sqrt{\eta N_*t+n_{\mathrm{pix}}(\eta N_bt+RN^2)}}
\end{equation}
where $t$ is the exposure time and $RN$ the readout noise of each
pixel.  $n_{\mathrm{pix}}$ and $\eta$ are the number of pixels
occupied by a single star and the system total efficiency,
respectively, which depend on the optical design and detector
technology. $N_*$ is the photon rate from the source and $N_b$ the
photon rate per pixel due to sky background.

For an observation carried out in a bandwidth of
$\Delta\lambda=\lambda_2-\lambda_1$, we have
\begin{equation}
  N_*=\frac{1}{16h}\bigg(\frac{d}{d_L}\bigg)^2\int_{\nu_1}^{\nu_2}L_\nu d\ln\nu
\end{equation}
for a given source with a specific luminosity $L_\nu$ and a luminosity distance of $d_L$, and
\begin{equation}
  N_b=\frac{1}{4}\pi d^2\int_{\Delta\Omega_p}\int_{\nu_1}^{\nu_2}R_\nu d\nu d\Omega
\end{equation}
where $R_\nu$ is the sky background photon flux per solid angle at
frequency $\nu$ and $\Delta\Omega_p$ the solid angle corresponding
to each pixel. The parameter $d$ is the diameter of each camera.
Based on the experience of \it GALEX\rm, the sky background
brightness is typically 27.5(26.5)$\mathrm{mag\ arcsec^{-2}}$, which
corresponds to a photon flux of $1\times10^{-6} (10^{-5})\
\mathrm{photons\ s^{-1}\ cm^{-2}\ arcsec^{-2}}$ in the FUV and NUV
bands, respectively (Martin et al. 2005). $\Delta\Omega_p$ can be
determined by $\Delta\Omega_p=(\alpha d_p)^2$, where
$\alpha=206265\arcsec/f$ and $f$ is the focal length, and $d_p$ is
the CCD pixel size.

For an observation carried out with both a given exposure time and a
fixed pixel scale, the limiting magnitude can be obtained from Eq.
(9) for a given signal-to-noise (S/N) ratio threshold. In the
limiting magnitude estimation, we only consider the limiting case in
which noise is dominated by both the sky background and readout
noise. We multiply a factor of 10 to the used background counter
rate to take into account of the noise contributed by potential
stray light, dark current, and flatfield correction.

\subsubsection{TDE}

The detection rate of TDEs occurring in the Universe at $z\ll1$ is
estimated by following Strubbe \& Quataert (2009)
\begin{equation}
  \tiny
  \Gamma_{\mathrm{TDE}}=\int_{\Delta\Omega}\int_0^{z_{\mathrm{max}}}\int_{M_{\mathrm{BH}}}\int_{R_p}^{R_T}\frac{dn}{d\ln M_{\mathrm{BH}}}
  \frac{d\gamma}{d\ln R_p}\frac{dV}{dzd\Omega}d\ln R_pd\ln M_{\mathrm{BH}}dzd\Omega
\end{equation}
where $\gamma$ is the tidal distribution rate for a single SMBH. We
ignore the dependence of $\gamma$ on $R_p$, and adopt the value of
$\gamma=10^{-5}\ \mathrm{yr^{-1}}$ reported in Donley et al. (2002),
which results in $d\gamma/d\ln R_p=\gamma/\ln(R_T/R_p)$.
$z_{\mathrm{max}}$ is the maximum redshift that is determined from
both the intrinsic luminosity and limiting magnitude calculated
above. We adopt a universal SMBH density
$n\simeq10^{-2}\mathrm{Mpc^{-3}}$ predicted in Hopkins et al.
(2007), although the density falls slightly at the high mass end.
$\Delta\Omega=N\delta\Omega$ is the total sky area that is
simultaneously surveyed by a set of $N$ cameras, and each camera has
an FoV of $\delta\Omega$.

The lower panel in Figure 3 shows the estimated detection rates
$\Gamma_{\mathrm{TDE}}$ in the cases with different $N$, as a
function of $\Delta\Omega$. The fiducial parameters used in the
predictions are listed in Table 2. One can see from the figure that
for a given $\Delta\Omega$, $\Gamma_{\mathrm{TDE}}$ generally
increases with $N$, which can be understood by a reduced
sky-background noise resulting from a decreasing value of
$\Delta\Omega_p$. In the case of a given $N$, although
$\Gamma_{\mathrm{TDE}}$ generally increases with $\Delta\Omega$, the
increment decreases evidently with $\Delta\Omega$. The decreased
increment results from an enhanced sky-background noise (i.e., an
enhanced $\Delta\Omega_p$) that neutralizes the increment caused by
an enhancement in total sky coverage.

\begin{figure}
\plotone{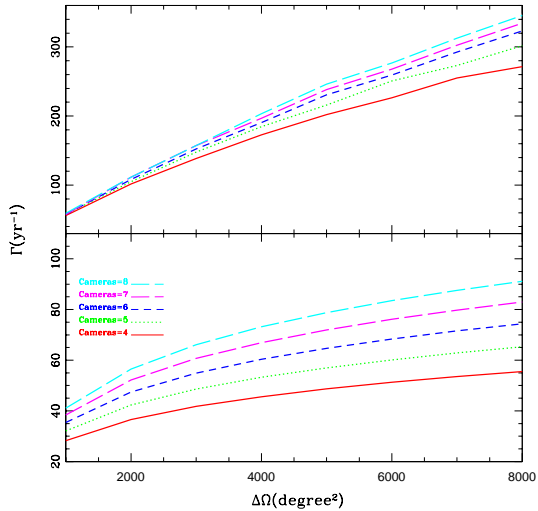} \caption{\it Lower panel: \rm the
predicted detection rates for the TDE cases plotted against the
total sky area covered by an NUV time-domain survey. The different
colors represent the predictions with different numbers of cameras.
\it Upper panel: \rm the same as the lower panel but for the SBO
cases. }
\end{figure}

\subsubsection{SBOs}

With the $z_{\mathrm{max}}$ determined from the ``CCD'' equation, we
estimate the detection rate of and SBO, $\Gamma_{\mathrm{SBO}}$
through the integral
\begin{equation}
  \Gamma_{\mathrm{SBO}}=\int_{\Delta\Omega}\int_0^{z_{\mathrm{max}}}f_{\mathrm{SNII}}R_{\mathrm{SN}}\frac{dV}{dzd\Omega}dzd\Omega
\end{equation}
where $f_{\mathrm{SNII}}=0.634$ is the fraction of type-II SNe in
core-collapse (CC) SNe (Li et al. 2011).
$R_{\mathrm{SN}}=k_{\mathrm{cc}}\psi(z)$, in which
$k_{\mathrm{cc}}=0.0068 M_\odot^{-1}$ is the CC-SNe conversion
efficiency per stellar mass based on the Salpter initial mass
function (IMF) with $m_{\mathrm{min}}=8M_\odot$ and
$m_{\mathrm{max}}=30M_\odot$ (e.g., Walmswell \& Eldridge 2012;
Tanaska et al. 2016). $\psi(z)$ is the cosmic star formation rate
(Madau \& Dickison 2014):
\begin{equation}
  \psi(z)=0.015\frac{(1+z)^{2.7}}{1+[(1+z)/2.9]^{5.6}}\ \mathrm{M_\odot\ yr^{-1}\ Mpc^{-3}}
\end{equation}

The predicted detection rates, which are similar to the TDE cases,
can be found in the upper panel of Figure 3.

\subsubsection{Super Flares of solar-like stars and M dwarfs}

Solar-like stars (i.e., type G and K dwarfs) and M dwarfs frequently
show super flares with releases of energy of $10^{33-36}$erg and
$10^{32-35}$erg, respectively. The energy is mainly radiated from
the corona and chromosphere in soft X-ray and UV bands. It makes the
flares more easily detected in an NUV time-domain survey since the
UV emission of the quiescent counterparts is much weaker than that
of the flares. We predict the detection rates of super flares for
solar-like stars and M dwarfs by the following estimation
\begin{equation}
  \Gamma_{\mathrm{flares}}=\int_{\Delta\Omega}\int_0^{r_{E_1}}\int_{m_1}^{m_2}4\pi r^2\rho_\star\int_{E_1}^{E_2}\frac{dn}{dm}\frac{dN}{dE}dEdrd\Omega dm
\end{equation}
where $dN/dE\propto E^{-\alpha}$ is the flare frequency distribution
function that is normalized by an occurrence rate of
$2.3\times10^{-3}\ \mathrm{flares\ yr^{-1}\ star^{-1}}$ (Shibayama
et al. 2013), and $dn/dm$ is the normalized Salpeter IMF. Previous
statistical studies show that $\alpha\sim2$ for both G and M dwarfs
(e.g., Maehara et al. 2012; Crosby et al. 1993; Aschwanden et al.
2000a,b; Schrijver et al. 2012; Shibayama et al. 2013; Yang et al.
2017). $r_{E_1}$,  which can be determined through the ``CCD''
equation, is the maximum distance detectable for a super flare
having a released energy of $E_1$. In our estimation, the stellar
density $\rho_\star$ for stars near the Sun is adopted to be 0.14
$\mathrm{star\ pc^{-3}}$ (Gregersen 2010).

Figure 4 shows the predicted detection rates for both type G and M
dwarfs as a function of sky area coverage. Compared to the TDE
cases, the effect on detection rate caused by the number of cameras
is slight since the background noise is reduced significantly due to
the short exposure (i.e., 30 seconds).

\begin{figure}
\plotone{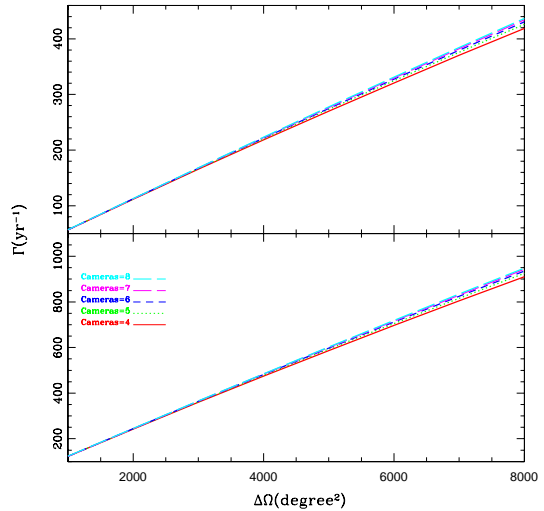} \caption{The same as in Figure 3 but for
flares occurring in G (the upper panel) and M (the lower panel)
dwarfs.}
\end{figure}

\section{Mission Concept}

By updating the concept proposed about the Chinese Space Station in
2011, here we propose a small satellite mission dedicated to an NUV
wide-angle time domain survey. A dedicated transient survey in the
NUV band is still rare. To our knowledge, there are only two
proposed missions in the works. One is the \it ULtraviolet TRansient
Astronomical SATellite (ULTRASAT) \rm (Sagiv et al. 2014), and the
other is the Wide-field Ultraviolet Imagers available for different
platforms (Mathew et al. 2018).

The instruments proposed by us are composed of a set of NUV
wide-angle cameras that are used for detecting transient events.
With the predictions described in the last section, the NUV
wide-angle camera array is proposed to contain eight cameras that
provide a total sky coverage of 3000$\mathrm{deg^2}$. The FoV and
focal length of each camera are $20\times20\mathrm{deg^2}$ and
260mm, respectively. Each of the cameras has a diameter of 20cm. An
NUV-enhanced 6k$\times$6k CCD with $\delta$-doping technology
(Nikzad et al. 2011) is equipped on each camera at the focal plane.
The pixel size of the CCD is 15$\mathrm{\mu m\ pixel^{-1}}$, which
yields a spatial resolution of $12\arcsec\ \mathrm{pixel^{-1}}$.

The cadence of the proposed survey ranges from 30s to 300s, which is
necessary for detecting star flares and SBOs. Dedicated on-board
software will be developed to detect transient events in real time
not only from single exposures, but also from combined images. After
a transient alert is generated, the alert and corresponding
sub-images centered on the transient should be transmitted to the
ground. The mission requires not only a suppression of the light
pollution resulting from stray light, but also a continuous,
unobstructed FoV of the celestial sphere. Both issues could be
addressed by a selection of a high Earth orbit. We suggest that the
low-cost eccentric ``P/2-HEO'' orbit (Gangestad et al. 2013) adopted
by the Transiting Exoplanet Survey Satellite (TESS) mission (Ricker
et al. 2015) is a good solution for both issues.
The mission is proposed to adopt an anti-solar pointing law, which enables ground facilities to follow-up the transients in
both imaging and spectroscopy as soon as possible.

\section{Conclusion}

By performing an analysis and simulations on the detection ability
in the NUV band for some major transient events (i.e., TDEs, SBOs
and flares of stars), here we propose an updated space-based patrol
mission dedicated to searching for NUV transient events. The
proposed mission stems from the original proposal submitted for the
Chinese Space Station mission. The updated mission is composed of a
set of eight small wide-field NUV cameras, each with a diameter of
20cm. The total sky area simultaneously covered by the NUV cameras
can be as large as 3000deg$^2$. The survey cadence is designed to
range from 30 to 300s.

%

\acknowledgments

We thank the anonymous referee for his/her careful
review and helpful suggestions that improved the manuscript.
The study is supported by the National Basic Research Program of
China (grant 2014CB845800), the NSFC under grants 11533003, and the
Strategic Pioneer Program on Space Science, Chinese Academy of
Sciences, Grant No. XDA15052600. JW is supported by the National
Natural Science Foundation of China under grants 11473036 and
11773036.



\end{document}